\documentclass[11pt]{article}
\usepackage{amsmath}
\usepackage{amssymb}
\usepackage{amsfonts}
\usepackage{fleqn} 
\usepackage{epsfig}
\usepackage{graphicx} 
\usepackage{mathrsfs}	

\date{\today}
\newcommand{\bmat}{\left(\begin{array}}
\newcommand{\emat}{\end{array}\right)}
\newcommand{\be}{\begin{equation}}
\newcommand{\ee}{\end{equation}}
\newcommand{\ba}{\begin{eqnarray}}
\newcommand{\ea}{\end{eqnarray}}

\def\lsim{\ensuremath{\lesssim}
}

\def\be{\beta}

\def\relic{\Omega_{\tilde{\chi}}}

\DeclareMathOperator{\re}{Re}

\DeclareMathOperator{\Tr}{Tr}

\begin{document}
\date{}
\title{
\begin{flushright}
\normalsize
CERN-PH-TH/2006-247\\
LPT--Orsay 06-78 \\
TUM-HEP-653/06 
\end{flushright}
\vskip 2cm
{\bf\huge Metastable Vacua in Flux Compactifications and Their Phenomenology
}\\[0.8cm]}

\author{{\bf\normalsize
Oleg Lebedev$^{1}$\!,
Val\'eri L\"owen$^2$\!,
Yann Mambrini$^3$\!,}\\
{\bf\normalsize 
Hans Peter Nilles$^2$\!,
Michael Ratz$^4$\!}\\[1cm]
{\it\normalsize
${}^1$ CERN, Theory Division, CH-1211 Geneva 23, Switzerland}\\[0.1cm]
{\it\normalsize
${}^2$ Physikalisches Institut der Universit\"at Bonn,}\\[-0.05cm]
{\it\normalsize Nussallee 12, 53115 Bonn,
Germany}\\[0.1cm]
{\it\normalsize
${}^3$ Laboratoire de Physique Th\'eorique,  
Universit\'e Paris-Sud, }\\[-0.05cm]
{\it\normalsize   F-91405 Orsay, France}\\[0.1cm]
{\it\normalsize
${}^4$ Physik Department T30, Technische Universit\"at M\"unchen,}\\[-0.05cm]
{\it\normalsize 85748 Garching,
Germany}
}
\maketitle \thispagestyle{empty} 
\begin{abstract}
{ 
In the context of flux compactifications,
metastable vacua with a small positive cosmological constant are obtained by
combining a sector where supersymmetry is broken dynamically with
the sector responsible for moduli stabilization, which  is known as the 
$F$--uplifting. We analyze  this procedure in a model--independent way
and   study phenomenological properties of the resulting  vacua.
}
\end{abstract}

\clearpage

\section{Introduction}

Recent progress in string theory compactifications with fluxes 
\cite{Giddings:2001yu}  has facilitated construction of string models with all
moduli stabilized, zero or small cosmological constant, and broken
supersymmetry. In the model of Kachru--Kallosh--Linde--Trivedi (KKLT)
\cite{Kachru:2003aw}, the complex structure moduli and the dilaton are
stabilized by fluxes on an internal manifold, while the K\"ahler ($T$) modulus 
is stabilized by non--perturbative effects such as gaugino condensation. The 
K\"ahler potential and the superpotential  for the $T$--modulus  are given by
\begin{equation}
 K~=~-3\ln (T + \overline{T})\;,\quad 
 W~=~ W_0 -A\,\mathrm{e}^{-a\, T} \;,
\end{equation}
where $W_0$, $A$ and $a$ are model-dependent constants. Minimization of the
corresponding scalar potential reveals that  supersymmetry is unbroken at the
minimum and  the vacuum energy  is negative and large in magnitude. To achieve a
small and positive cosmological constant in this setup, KKLT introduced an
anti--D3 brane whose contribution to vacuum energy can be adjusted arbitrarily. 
However, such a contribution breaks supersymmetry explicitly. It was later
suggested in \cite{Burgess:2003ic}  that a similar uplifting effect could  be
achieved in the framework of spontaneously broken SUSY by including the
$D$--terms. This procedure however cannot uplift the KKLT  minimum due to the
supergravity relation $D \propto  F =0$ \cite{Choi:2005ge}.  It can {\it
potentially} be used to uplift non--SUSY minima such as those at exponentially
large compactification volume \cite{Balasubramanian:2005zx,Conlon:2006wz}.

One of the obstacles to realizing the simplest KKLT scenario in supergravity
is posed by the no--go theorem of Refs.~\cite{Brustein:2004xn,Gomez-Reino:2006dk,Lebedev:2006qq}
\footnote{A generalization of this theorem can be found in \cite{Lust:2006zg}.}.
It states that 
\begin{center}
\begin{minipage}{0.95\textwidth}
{\it if the modulus $T$  is the only light field and
its   K\"ahler potential  is   $K=-n\,\ln (T + \overline{T})$ ($1\le n\le3$), 
de Sitter (dS) or Minkowski vacua with broken supersymmetry are not possible for
any superpotential. }
\end{minipage}
\end{center}
Thus, it is necessary to include additional fields in the system (or modify the
K\"ahler potential \cite{deAlwis:2005tf}) providing  the goldstino  which is
necessary to make the gravitino massive.  In this case, dS or Minkowski vacua 
with  spontaneously  broken supersymmetry  can be  obtained  due to the
$F$--terms of hidden matter fields \cite{Lebedev:2006qq} (a somewhat similar
approach was considered in \cite{Saltman:2004sn}).  Since matter fields are as
generic  as moduli in string constructions, this provides an interesting
alternative to common scenarios with moduli/dilaton  dominated SUSY breaking.
In this article, we will   follow   our earlier work  (LNR)
\cite{Lebedev:2006qq}.

Interest in this approach has been bolstered by recent work of Intriligator, Seiberg and
Shih  (ISS)  \cite{Intriligator:2006dd} on dynamical SUSY breaking in metastable vacua.
They have found that metastable vacua with broken supersymmetry are generic and 
realized even in simple systems like SUSY QCD. These vacua are long--lived 
and can be combined with the KKLT sector to achieve a small cosmological constant
and acceptable supersymmetry breaking \cite{Dudas:2006gr,Abe:2006xp,Kallosh:2006dv}.

In this work, we analyze the $F$--term uplifting  of the KKLT minimum in a
model--independent way and study SUSY phenomenology of the resulting vacua. 
Before we proceed, let us give a few relevant supergravity formulae. The
supergravity scalar potential is given by \cite{Cremmer:1982en}
\begin{equation}
 V~=~ \mathrm{e}^G\, (G_i\, G_{\bar\jmath}\, G^{i \bar\jmath} -3)   +   
 {1\over 2}\,\re (f_a)^{-1} \,D^a D^a   \;,
\label{V}
\end{equation}
where $G=K +\ln \vert W \vert^2$ with  $K$ and $W$ being the K\"ahler potential
and the superpotential, respectively; $f_a$ is the gauge kinetic function, and
$D^a$ are the $D$--terms. A subscript $i$  denotes differentiation with respect
to the $i$-th field. $G^{i \bar\jmath}$ is the inverse K\"ahler metric.
The gravitino mass is given by
\begin{equation}
 m_{3/2}~=~ \mathrm{e}^{G/2} \;,
\end{equation}
and the SUSY--breaking $F$--terms are
\begin{equation} 
 F^i~=~\mathrm{e}^{G/2}\, G^{i \bar \jmath}\, G_{\bar\jmath} \;.
\end{equation}

In what follows, we first review problems with the $D$--term uplifting and then 
focus on the uplifting with the  $F$--terms.

\section{Problems with the $\boldsymbol{D}$--term uplifting}

There are two problems with the $D$--term uplifting scenario. First,
supersymmetric minima cannot be uplifted by the $D$--terms \cite{Choi:2005ge}.
The reason is that, in  supergravity  \cite{Cremmer:1982en},
\begin{equation}
 D^a  \propto {1\over W}\, D_i W \;, 
\end{equation}
where $D_i W \equiv \partial_i W +  W \partial_i K$. In supersymmetric
configurations, $\langle D_i W \rangle =0$ and the $D$--terms vanish (unless
$W=0$). Thus,  only non--supersymmetric minima can potentially be uplifted.

Second, the $D$--term uplifting of non--supersymmetric vacua
does not work either if the gravitino mass is hierarchically
small \cite{Choi:2006bh} (unless moduli are exponentially large). 
Indeed,  for matter on D7 branes,  the gauge kinetic
function is given by
\begin{equation}
f~=~T \;,
\end{equation}
and the $D$--term of an anomalous U(1) is
\begin{equation}
 D~\propto~{E \over  \re T} + \sum_i q_i\, \vert \phi_i \vert^2 \;,
\end{equation}
where $E$ is a constant related to the trace of the anomalous U(1) and $\phi_i$
are VEVs of fields carrying anomalous charges $q_i$.  At the minimum of the
scalar potential,
\begin{equation}
 V_T~=~0 \;,
\end{equation}
which from Eq.~(\ref{V}) implies symbolically
\begin{equation}
 m_{3/2}^2 + D^2 + D~=~0  \;.
\end{equation}
Here we have neglected all coefficients and assumed that there are no
very large ($10^{15}$)  or very small factors in this equation.
Using  $m_{3/2} \sim 10^{-15}$ (in Planck units) as favoured  by phenomenology,
this equation is solved by
\begin{equation}
 D~\sim~  m_{3/2}^2 ~\ll~m_{3/2}~ \sim~ F \;. 
\end{equation}
Thus, the $D$--term is much smaller than the $F$--terms and
$D^2 \sim m_{3/2}^4 $ cannot uplift an AdS minimum with $V_0 \sim - m_{3/2}^2 $
to zero vacuum energy.
This mechanism can only work for  a heavy gravitino, e.g.\ $m_{3/2} \sim 1$.
The existing examples of the $D$--term uplifting confirm this conclusion
\cite{Villadoro:2005yq,Achucarro:2006zf,Parameswaran:2006jh,Dudas:2006vc}
(for related work, see also \cite{Brummer:2006dg,Haack:2006cy,Braun:2006se}).

\section{$\boldsymbol{F}$--term uplifting}

It has been shown by LNR \cite{Lebedev:2006qq} that
the $F$--term uplifting mechanism is viable and works for a hierarchically small
gravitino mass.   
The $F$--uplifting   in its simplest form amounts to  combining  a sector
where supersymmetry is broken spontaneously in a metastable  dS vacuum  with the KKLT
sector.   Since the $T$--modulus is heavy, the resulting minimum of the system
is given approximately by the minima in the separate subsectors.  Then $T$ gives
only a small contribution to SUSY breaking and  the cosmological constant can be
adjusted to be arbitrarily small.  Let us study this procedure in more detail.

\subsection{SUSY breaking sector}

Consider a (hidden  sector) matter field $\phi$ with 
\begin{equation}
 K~=~ \overline{\phi} \phi \;,\quad W~=~ \mathcal{W}(\phi) \;.
\end{equation}
Suppose for simplicity that the minimum of the corresponding scalar potential 
\begin{equation}
 V~=~\mathrm{e}^G\, (G_\phi\, G_{\overline{\phi}}  -3) \;,
\end{equation}
is at real $\phi$.
The non--supersymmetric minimum is found from
\begin{equation}
 V_\phi~\propto~G_\phi^2 +G_{\phi \phi} -2~=~0 \;.
\end{equation}
Denoting  this minimum by $\phi_0$, supersymmetry is broken by
$F^\phi ~\sim~ \vert \mathcal{W} \vert\, G_{\phi} \bigl\vert_{\phi_0}$. 
The vacuum energy can be chosen to be positive,
\begin{equation}
 V(\phi_0)~>~ 0 \;,
\label{V0}
\end{equation}
and arbitrarily small by adjusting $\mathcal{W}(\phi)$.
In this case, $F^\phi \sim \vert \mathcal{W} 
(\phi_0) \vert$ and, assuming that the potential is not very steep,  
the mass of $\phi$ is typically   of order $ \vert
\mathcal{W} (\phi_0) \vert $.

\subsection{KKLT sector}

This sector consists of the $T$--modulus with the  usual K\"ahler potential and
the superpotential  induced by fluxes and  gaugino condensation,
\begin{equation}
 K~=~-3\ln (T + \overline{T}) \;,\quad 
 W~=~\mathscr{W}(T)~\equiv~ W_0 -A \mathrm{e}^{-a\, T} \;,
\end{equation}
with $A\sim 1$, $a\gg 1$. 
If the observable matter is placed on D7 branes, the SM gauge couplings require
$\re T \simeq 2$ at the minimum.

The scalar potential is
\begin{equation}
 V~=~\mathrm{e}^G \,( G_T\, G_{\overline{T}}\, G^{T \overline{T}} -3 ) \;,
\end{equation}
and its  SUSY minimum is determined by
\begin{equation}
 G_T~=~0 \;.
\end{equation}
The solution is
\begin{equation}
 T_0~\approx~ -{1\over a}\, \ln {W_0 \over a} \;,
\label{kklt}
\end{equation}
where again we have taken $T$ to be real. The  corresponding vacuum energy 
is given by
\begin{equation}
 V(T_0) ~=~-3\,\mathrm{e}^G \sim - \left\vert \mathscr{W}(T_0) \right\vert^2 \;.
\end{equation}

\subsection{KKLT + SUSY breaking sector}

Now we combine the two sectors. The full K\"ahler potential and the superpotential
are given by\footnote{This setup can be realized for matter on D7 branes. The corresponding
K\"ahler metric can be found in  \cite{Lust:2004dn}.
Following  KKLT, here we assume that
the dilaton and complex structure moduli have been integrated out and
neglect possible corrections to the K\"ahler potential \cite{deAlwis:2005tf}
 due to this  procedure.} 
\begin{eqnarray}
 K & = & \vert \phi \vert^2 -3 \ln (T + \overline{T})   \nonumber \;,\\
 W & = &  \mathcal{W}(\phi) + \mathscr{W}(T)  \;.
\end{eqnarray}
The question is now how much the minimum of the system deviates from the minima
of the separate subsectors.

Consider the system in  the vicinity of the reference point $(\phi_0, T_0)$.
At $T=T_0$, the superpotential for $\phi$ is
\begin{equation}
 W~=~ \mathcal{W}(\phi) + \mathscr{W}(T_0) \;.
\end{equation}
Similarly, at $\phi=\phi_0$ the superpotential for $T$ is 
\begin{equation}
 W~=~ \mathscr{W}(T) + \mathcal{W}(\phi_0) \;.
\end{equation}
Thus, the constant  terms in the superpotential  shift relative  to those of 
the original subsectors. 
It makes sense to compare the true minimum of the system to the minima of the
subsectors with shifted superpotentials.  For example, $T_0$ should be  defined
as  the minimum of the KKLT subsector with the superpotential $  \mathscr{W}(T)
+ \mathcal{W}(\phi_0)  $ and similarly for the  $\phi$ subsector. This can be
done iteratively.

The total potential is now
\begin{equation}
 V~=~\mathrm{e}^G\, ( G_\phi\, G_{\overline{\phi}} 
 + G_T\, G_{\overline{T}}\, G^{T \overline{T}} -3) \;.
\end{equation}
Let us see if $(\phi_0, T_0)$ is a stationary point. We have  
\begin{eqnarray}
 V_\phi& =& G_\phi\, V +  \mathrm{e}^G\,
 {\partial \over \partial \phi }( G_\phi G_{\overline{\phi}})
 +\mathrm{e}^G\,{\partial \over \partial \phi }( G_T\, G_{\overline{T}}\, 
 G^{T \overline{T}} ) \;, \nonumber\\
 V_T& =& G_T\, V + \mathrm{e}^G\,{\partial \over \partial T }
 ( G_T\, G_{\overline{T}}\, G^{T \overline{T}} ) +
 \mathrm{e}^G\,{\partial \over \partial T }( G_\phi\, G_{\overline{\phi}}) \;.
\end{eqnarray}
Consider $V_\phi$. It is zero because the first two terms represent the
equations of motion for the separate $\phi$--subsector, and the third term is
proportional to $G_T$ which is zero at $T_0$. 
Consider now $V_T$. The first two terms are zero due to $G_T(T_0)=0$. The last
term  however does not vanish,
\begin{equation}
 \mathrm{e}^G\,{\partial \over \partial T }( G_\phi G_{\overline{\phi}})
 ~\sim~   
  m_{3/2}^2 \;,
\end{equation}
where we have used $G_\phi, W_T/W \sim 1$ at $(\phi_0,T_0)$.
It is non--zero but small compared to $V_{T\bar T} \sim a^2 m_{3/2}^2$. 
Therefore, the (heavy)  modulus  shifts  slightly from $T_0$. 
Finally, the vacuum energy at $(\phi_0, T_0)$ equals that of the $\phi$--subsector
from Eq.~(\ref{V0}).

We see that the stationary point conditions are ``almost'' satisfied at 
$(\phi_0, T_0)$. Let us now compute how much the true minimum is shifted
compared to  $(\phi_0, T_0)$. Suppose the true minimum is at $\phi_0 + \delta
\phi, T_0 +\delta T$. At this point,
\begin{eqnarray}
 V_\phi (\phi_0 + \delta \phi,T_0 +\delta T) & =& 0 \;, \nonumber\\
 V_T (\phi_0 + \delta \phi,T_0 +\delta T) & =& 0 \;.
\end{eqnarray}
The $(\phi,T)$ system has been studied  in detail in LNR    \cite{Lebedev:2006qq},
while here we will, 
for simplicity,  treat $T$ and $\phi$ as real 
variables
and expand this to first order in $\delta \phi, \delta T$,
\begin{eqnarray}
 V_{\phi \phi}\, \delta \phi +  V_{\phi T}\, \delta T    & = & 0 \;, \nonumber\\
 V_T  +  V_{ T T}\, \delta T + V_{ T \phi }\, \delta \phi & = & 0 \;,
\end{eqnarray}
where we have used $V_\phi (  \phi_0, T_0) =0$ as explained above. 
The solution is
\begin{eqnarray}
 \delta T & = &  {V_T \over V_{T\phi}^2/V_{\phi \phi} - V_{TT}} \;, \nonumber\\
  \delta \phi & = & - {V_{T \phi} \over V_{\phi \phi}}\, \delta T\;.
\end{eqnarray}
In the large $a$ limit, $\delta T \sim 1/a^2$ and $\delta \phi \sim 1/a~$.\footnote{The relation
$\delta T \sim \delta \phi / a $ can also be understood from rescaling the variable $T$,
$T'=aT$, which only affects the overall normalization of $V$ and implies $\delta T' \sim \delta \phi  $. }

Supersymmetry is now broken by $F^\phi$ and $F^T$ with the latter
giving a small contribution,
\begin{equation} 
 F^T~\sim~ \mathrm{e}^{G/2}\,{W_{TT} \over W}\,\delta T
 ~ \sim~ {1\over a}\, m_{3/2} \;. 
\end{equation}
Finally, the cosmological constant can be chosen to be arbitrarily small 
by adjusting  parameters of the $\phi$--subsector, i.e.\ ${\cal W}(\phi)$.

\subsection{Example}

As a simple example, consider a combination of the KKLT and the Polonyi model
\cite{Polonyi:1977}.
The superpotential of the Polonyi model is given by
\begin{eqnarray}
 \mathcal{W}(\phi) ~=~ c +\mu^2\, \phi \;,
\end{eqnarray}
where $c$ and $\mu^2$ are constants.
A non--supersymmetric Polonyi  vacuum is determined by 
\begin{equation}
 G_\phi^2 +G_{\phi \phi} -2 ~=~0\;.
\end{equation}
Choosing 
\begin{equation}
G_\phi^2 ~=~ 3+\epsilon \;,
\end{equation}
with $\epsilon \ll 1$,  the vacuum energy is 
\begin{equation}
V_0 ~\sim~ \epsilon\, \mu^4 \;.
\end{equation}
The solution to first order in $\epsilon$ is given by 
\begin{equation}
c ~\approx~ \mu^2\, \Bigl( 2- \sqrt{3} - {\sqrt{3} \over 6} ~\epsilon \Bigr) \;,
\quad\phi~\approx~\sqrt{3}-1 + {  \sqrt{3}-3 \over 6}\, \epsilon \;.
\end{equation}
The mass of the Polonyi field is set by $\mu^2$.

\begin{figure}[!h!]
\centerline{\includegraphics{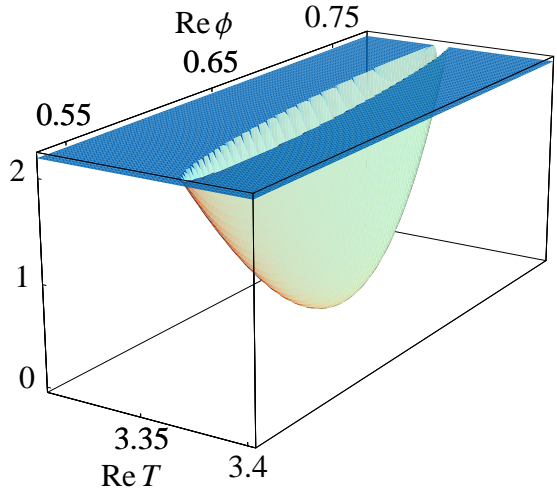}}
\caption{
Scalar potential   of the KKLT + Polonyi model
(in units of $10^4/\mu^4$).
 \label{Fplot}}
\end{figure}

This system can be used to uplift the AdS minimum of KKLT as explained above. 
Since $\mu^2 \sim W_0$, the modulus is
heavy compared to the Polonyi field. As a result, it shifts  only slightly from
the original position and its contribution to SUSY breaking is suppressed.
The resulting vacuum energy can be made arbitrarily small by adjusting $\epsilon$
and without affecting other aspects of the system.

The scalar potential for $A=1$, $a= 12$, $\mu = 10^{-8}$ is displayed in
Fig.~\ref{Fplot}.

\subsection{Relation to ISS}

An  interesting class  of  matter sectors with dynamically broken supersymmetry
is provided by ISS \cite{Intriligator:2006dd}. In this case, small $W_0$ is generated dynamically through
dimensional transmutation.  The ISS examples include SUSY QCD with massive
flavours, whose dual is described by the superpotential
\begin{equation} 
 W~=~ h\,\Tr \phi\, \Phi\, \widetilde{\phi} - h\, \mu^2\,\Tr\Phi \;.
\end{equation}
Here $\phi_i^a$, $\widetilde{\phi}_a^{\bar \jmath}$, $\Phi^i_{\bar \jmath}$ are the quark and
meson fields  with $1\le i,j\le N_f$ and $1\le a\le N$ being the flavour and
colour indices, respectively.  $h$ and $\mu^2$ are (dynamically generated)
constants.

This system possesses metastable vacua with broken supersymmetry and 
\begin{equation}
 V_0~=~ (N_f-N)\,\left\vert h^2\, \mu^4 \right\vert \;.
\end{equation}
Such vacua can be used for uplifting the KKLT minimum along the lines
described above. For more details, see \cite{Dudas:2006gr,Abe:2006xp,Kallosh:2006dv}.

\subsection{Remark on other schemes}

Although we have focused  our discussion on uplifting the KKLT minimum,
 it is clear that very similar considerations apply to other schemes.
Analogous systems arise in the heterotic string, with the
substitution $ K=-3\ln (T + \bar T) \rightarrow -\ln (S + \bar S)$
and $f=T \rightarrow S$. The analysis of SUSY breaking
can be carried out in a similar fashion with the same qualitative
conclusions.

A related analysis for M theory compactifications on $G_2$ manifolds
is given in \cite{Acharya:2006ia}.

\section{Soft terms}

The resulting pattern of the soft terms is a version of the
``matter dominated SUSY breaking'' scenario  \cite{Lebedev:2006qq}.
$F$--term uplifting  generally predicts heavy scalars with masses
of order the gravitino mass and light gauginos,
\begin{equation}
m_{1/2} ~ \ll ~m_0~ \sim~ m_{3/2}  \;.
\end{equation}
The suppression of the gaugino masses comes from the fact that the gauge
kinetic  function is independent of $\phi$ to leading order.

Let us now focus on the case $\langle \phi \rangle \ll 1$. Allowing for  the
K\"ahler potential coupling between $\phi$ and observable fields $Q_i$,
\begin{equation}
 K~=~ -3 \ln (T + \overline{T}) +  \overline{\phi} \phi  
 + \overline{Q}_i Q_i\, (T +\overline{T})^{n_i}
 \Bigl[         1+ \xi_i\, \overline{\phi} \phi + {\cal O}(\phi^4) \Bigr] \;, 
 \label{eq:Ksoft}
\end{equation}
where $n_i$ are effective ``modular weights'', we have \cite{Lebedev:2006qq}
\begin{eqnarray}
 M_a & =&  M_s\, \left[   \alpha_\mathrm{FLM} + b_a\, g_a^2   \right] \;, \\
 A_{ijk}& = &{}   - M_s\, \left[   3\,\alpha_\mathrm{FLM} -\gamma_i -\gamma_j -\gamma_k   \right] \;, \\
 m_i^2 & = & (16 \pi^2\, M_s)^2\, \left[    1-3\, \xi_i   \right] \;,
\end{eqnarray}
with $M_s \equiv m_{3/2}/16 \pi^2 $.
Here we follow the notation of \cite{Falkowski:2005ck}. $b_a$ and $\gamma_i$ are
the beta--function coefficients and the anomalous dimensions, respectively. 
The gaugino masses and A--terms receive comparable contributions from the modulus and the anomaly 
\cite{Randall:1998uk,Giudice:1998xp} as in Refs.~\cite{Choi:2005ge,Choi:2004sx},
while the scalar masses are dominated by the $F$--term of the uplifting field
$\phi$. The parameter $ \alpha_\mathrm{FLM}$ of \cite{Falkowski:2005ck} controls
the balance between the modulus and the anomaly contributions to $M_a$ and
$A_{ijk}$: at large $ \alpha_\mathrm{FLM}$ the modulus contribution dominates,
while  at small $ \alpha_\mathrm{FLM}$ the anomaly provides the dominant
contribution.

The modular weights have little effect on the soft terms as they only affect the
A--terms. Thus we have set them to zero. The important variables for
phenomenology are  $ \alpha_\mathrm{FLM}$ and $\xi_i$. 

An interesting  feature of the soft terms is that the gaugino masses unify at a
scale between the electroweak and the GUT scales \cite{Choi:2005uz} although
there is no new physics appearing there. This is true in models where the
non--universality in gaugino masses is given by the corresponding
beta--functions. In general,  loop--suppressed contributions to the gaugino
masses come also from the K\"ahler anomalies \cite{Bagger:1999rd,Binetruy:2000md} 
and string
threshold corrections \cite{Dixon:1990pc}. When these are suppressed (e.g.\ when
$ \langle \phi \rangle \ll 1$), the ``mirage unification'' occurs. This,
however, does not usually apply to the scalar masses.

In what follows,  we study SUSY phenomenology of models with the pattern of 
soft terms given above.

\section{Phenomenology}

The variable  string/GUT scale  parameters in our scheme are
$$ m_{3/2}~~,~~ \alpha_\mathrm{FLM}~~,~~ \xi_i~~,~~\tan\beta \;,$$
while, for simplicity, we fix the sign of $\mu$ to be positive.
Here $\xi_i$ can be different for Higgses and sfermions, but we
assume it to be   generation--independent.
Further, we restrict ourselves to  the range $0< \alpha_\mathrm{FLM} <30$
and $0 \leq \xi_i < 1/3 $.

 The ``matter domination'' scheme 
has distinct  phenomenology.
Compared to  ``mirage mediation''  extensively 
studied in  Refs.~\cite{Choi:2005ge,Endo:2005uy,Choi:2005uz,Falkowski:2005ck,%
Choi:2005hd,Loaiza-Brito:2005fa,Baer:2006tb},
our scenario differs in the scalar masses, which are now large and comparable
to the gravitino mass. The controlled non--universality in the gaugino 
masses and the A--terms makes it different from  mSUGRA and its extensions
with non--universal Higgs masses. 
We find that there are considerable regions of parameter space where 
the scheme is consistent with all phenomenological constraints.

\subsection{Constraints and observables}

Certain regions of  parameter space are excluded by absence of  electroweak symmetry breaking and
a charged/coloured LSP.
Among other constraints,
the most important ones come from the Higgs and chargino searches,
\begin{equation}
m_h ~>~114\,\mathrm{GeV} \;,\quad m_{\tilde \chi^+} ~>~103 \, \mathrm{GeV} \;.
\end{equation}
Due to heavy scalars in our scenario, we expect the lightest Higgs to be very similar to
the SM Higgs, hence the LEP constraint  $m_h >114\,\mathrm{GeV}$ applies. 
We further impose the $b \rightarrow s \gamma$ constraint from the B--factories \cite{cleo,belle},
$ 2.33 \times 10^{-4} \leq \mathrm{BR}(b \rightarrow s \gamma) \leq 4.15 \times 10^{-4}$.

We also take into account the dark matter constraint. That is, we assume that
the LSP is stable, has  thermal abundance and constitutes the dominant component
of dark matter. Then we impose the $3\,\sigma$ WMAP constraint on dark matter
abundance   $ 0.087\lsim\relic\, h^2\lsim 0.138  $ \cite{Spergel:2006hy}  and
exclude parts of parameter space.  We also display parameter space allowed by a
conservative bound $ 0.03  < \relic h^2 < 0.3  $. Note that the above 
assumptions may be relaxed which would open up more available parameter space.
For instance, the LSP abundance may be non--thermal or the LSP may only
constitute a small component of dark matter.

Having singled out  favoured regions of parameter space, we consider prospects of 
direct and indirect dark matter detection. Dark matter can be observed (``directly'')
via elastic scattering on target nuclei with nuclear recoil (see \cite{Munoz:2003gx,Bertone:2004pz}). 
This process is dominated
by the $Z$ and Higgs exchange. Indirect dark matter detection  amounts to observing
a gamma--ray flux from the Galactic center, which can be produced by  dark matter 
annihilation \cite{Prada:2004pi,Profumo:2005xd}.

In our numerical analysis, we use the public codes SUSPECT \cite{Suspect}, SOFTSUSY \cite{Allanach:2001kg},
DarkSUSY \cite{darksusynew} and
MicrOMEGAs \cite{micromegas}.

\subsection{Parameter space analysis}

\begin{figure}[h!]
\vspace*{-.4cm}
    \begin{center}
	\hskip -.3cm
       \epsfig{file=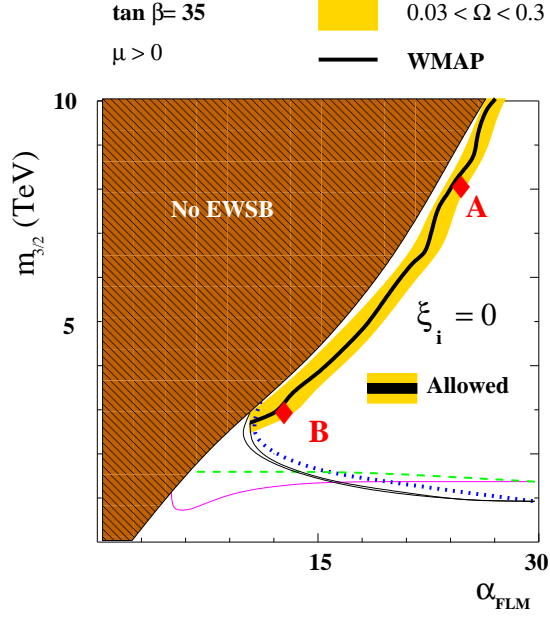,width=0.45\textwidth}
	\vskip -0.1cm
\caption{{\footnotesize
Allowed parameter space 
for tan$\beta=35$, $\xi_i=0$  and $\mu>0$.
The region below the light grey (green) dashed line
is excluded by the Higgs mass bound. The region below the 
the dotted line is excluded by  the chargino mass bound, while that 
below the solid (magenta) line is excluded by BR($b\to s\gamma$).
The narrow area between  the black  contours satisfies the 
 $3 \sigma$   WMAP constraint:
$0.087 \leq \Omega_{\tilde{\chi}_1^0}h^2 \leq 0.138$, 
whereas the yellow region satisfies 
$0.03 \leq \Omega_{\tilde{\chi}_1^0}h^2 \leq 0.3$.} }
        \label{fig:scantb35}
    \end{center}
\vspace*{-.5cm}
\end{figure}

We start with the case $\xi_i=0$ and $\tan\beta=35$. 
The allowed parameter space is shown in Fig.~\ref{fig:scantb35}, in yellow.
The chargino and the Higgs 
mass constraints require $m_{3/2}$ to be above a few TeV.
A large region is excluded due to no electroweak symmetry breaking (EWSB).
This can be understood by writing the EWSB condition 
in terms of the GUT input parameters. At $\tan\beta=5$,  
we have \cite{Kane:2002ap}
\begin{equation}
M_Z^2~=~-1.8\, \mu^2   -1.2\, m_{H_u}^2   +   5.9\, M_3^2 + 1.6\, m_{\tilde q_3}^2
+\dots\;, 
\label{mz}
\end{equation}
where $m_{\tilde q_3} $ is the third generation squark mass parameter.
In the case of heavy scalars, the dominant contribution 
is given by
$-1.2 m_{H_u}^2 + 1.6 m_{\tilde q_3}^2$, which must be positive.
The coefficient of $ m_{\tilde q_3}^2$ decreases with $\tan\beta$
due to the sbottom loops and at a certain critical value
electroweak symmetry remains unbroken. Thus, at low $\tan\beta$ 
 more parameter space is available.

Similarly, if we decrease the input value of $m_{\tilde q_3}^2$,  
electroweak symmetry gets restored. This means that
increasing $\xi_\mathrm{sf}$ widens the region excluded
by the ``no EWSB'' constraint.

The yellow region of Fig.~\ref{fig:scantb35} is favoured by dark matter
considerations. There the LSP is a mixed higgsino--bino and 
the correct relic density is achieved due to neutralino annihilation
and $\tilde{\chi}_1^0 \tilde{\chi}_1^+$, $\tilde{\chi}_1^0 \tilde{\chi}_2^0$ coannihilation
processes. The sample spectra for this region are given in Table~\ref{tab}.
To the left of the yellow region, the LSP relic density is below 0.03.
This part of parameter space is also viable if the LSP constitutes
only a fraction of dark matter. To the right of the yellow region,
the relic density is too large. In principle, this could also be consistent
with cosmological constraints if the dark matter production is non--thermal.

\begin{figure}[h!]
\vspace*{-.4cm}
    \begin{center}
	\hskip -.3cm
       \epsfig{file=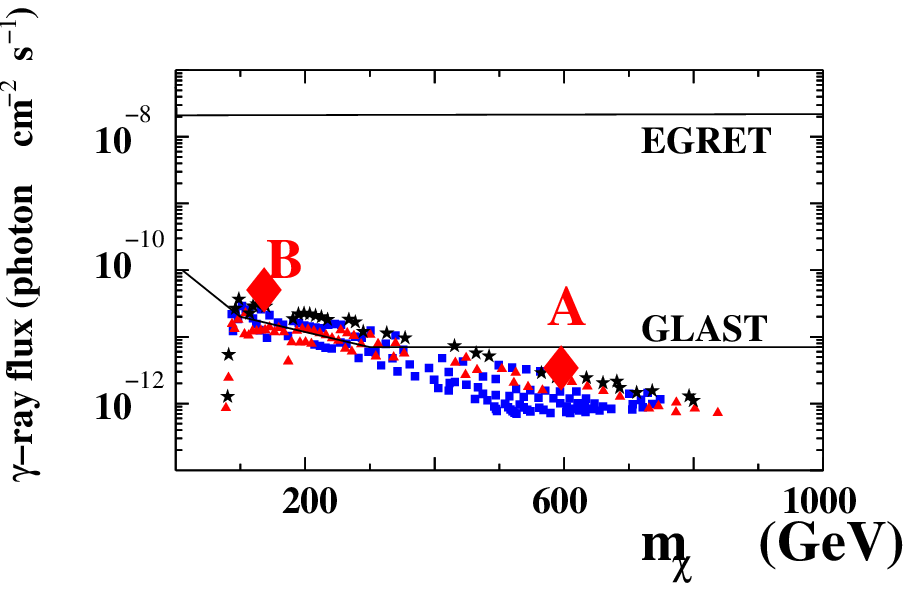,width=0.5\textwidth}
       \epsfig{file=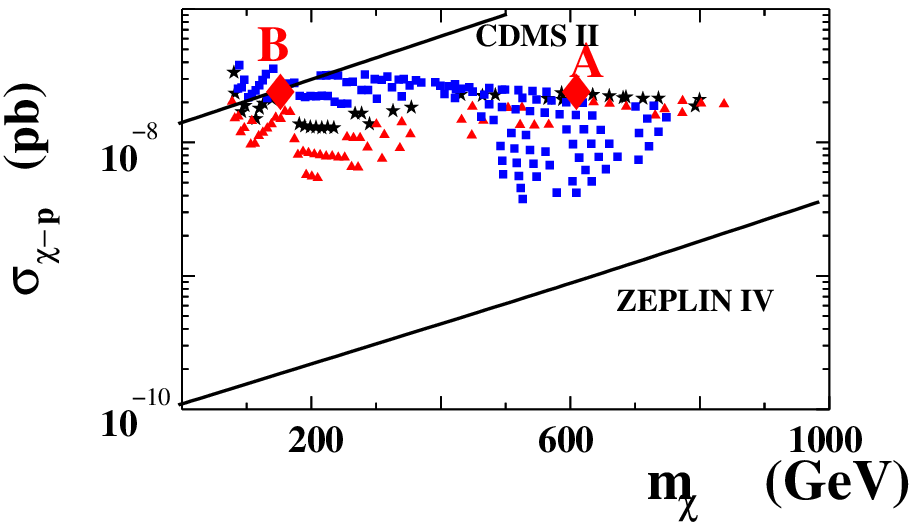,width=0.5\textwidth}

	\vskip -0.1cm
\caption{{\footnotesize 
Scatter plot of the gamma-ray flux $\Phi_{\gamma}$ 
for a threshold of 1 GeV (left)
and  the spin--independent neutralino--proton cross section (right)
as a function of the lightest neutralino mass $m_{\tilde{\chi}}$ 
for tan$\beta=35$, $\xi_i=0$ and $\mu>0$.
An   NFW dark matter profile with
with $\Delta \Omega \sim 10^{-5}$ sr is used.
All points in the figure satisfy the experimental bounds.
The light grey (red) triangles correspond to
$0.138<\Omega_{\tilde{\chi}^0_1}h^2<0.3$, 
black stars: 
$0.087<\Omega_{\tilde{\chi}^0_1} h^2<0.138$,
dark grey (blue) boxes:
$0.03<\Omega_{\tilde{\chi}^0_1} h^2<0.087$.
The solid lines represent the $5\sigma$ sensitivity  of the satellites.
} }
        \label{fig:detectb35}
    \end{center}
\vspace*{-.5cm}
\end{figure}

Prospects for indirect and direct detection of dark matter are presented in 
Fig.\ref{fig:detectb35}. 
Concerning the former,
the gamma ray flux from the Galactic center
is produced  in this case   
by an  s--channel $Z$ exchange or t--channel $\tilde{\chi}^+_1, \tilde{\chi}_2^0$
exchange. We see that relatively light neutralinos, 
$m_{\tilde{\chi}^0_1} \lsim 300$ GeV,  can be detected by GLAST,
a satellite based experiment to be launched in 2007, but are
beyond the reach of EGRET.
The neutralinos can also be detected directly via elastic scattering
on nuclei dominated by the t--channel $Z$ and Higgs exchange.
CDMS II (2007)  will probe part of the parameter space,  while ZEPLIN IV
 (2010) will cover the entire region allowed by WMAP. 
The scattering cross section is significant mainly due to  the $Z$--exchange
contribution and the higgsino--bino nature of the neutralino.

\begin{center} 
\begin{table}[h]
\centerline{
\begin{tabular}{|c|ccc|} 
\hline  
&\bf{A}&\bf{B}& \\ 
\hline  
$\tan \beta$ & 35 & 35 &  \\ 
$\alpha$& 23.8 & 11.9 &   \\ 
$m_{3/2}$ (TeV)& 8 & 3 &   \\
\hline 
$M_1$ & 625 & 125 &   \\
$M_2$ & 999 & 187 &   \\
$M_3$ & 2267 & 430 &   \\
\hline 
$m_{\tilde{\chi}^0_1}$ & 594 & 112 &   \\
$m_{\tilde{\chi}^0_2}$ & 635 & 159 &   \\
$m_{\tilde{\chi}^+_1}$ & 627 & 151 &   \\
$m_{\tilde g}$ & 2810 & 612 &   \\
\hline
$m_{h}$ & 127.1 & 121.1 &   \\
$m_{A}$ & 5972 & 2236 &   \\
$m_{H}$ & 5972 & 2236 &   \\
$\mu$   & 617 & 194 &   \\
\hline
$m_{\tilde t_1}$ & 4483 & 1732 &   \\
$m_{\tilde t_2}$ & 5477 & 2239 &   \\
$m_{\tilde c_1}, ~ m_{\tilde u_1}$ & 8171 & 2293 &   \\
$m_{\tilde c_2}, ~ m_{\tilde u_2}$ & 8172 & 2989 &   \\
\hline
$m_{\tilde b_1}$ & 6240 & 2241 &   \\
$m_{\tilde b_2}$ & 7249 & 2647 &   \\
$m_{\tilde s_1}, ~ m_{\tilde d_1}$ & 8170 & 2984 &   \\
$m_{\tilde s_2}, ~ m_{\tilde d_2}$ & 8172 & 2994 &   \\
\hline
$m_{\tilde \tau_1}$ & 7098 & 2657 &   \\
$m_{\tilde \tau_2}$ & 7568 & 2825 &   \\
$m_{\tilde \mu_1}, ~ m_{\tilde e_1}$ & 8001& 2989 &   \\
$m_{\tilde \mu_2}, ~ m_{\tilde e_2}$ & 8003 & 2996 &   \\
$m_{\tilde \nu_3}$& 8003 & 2988 &   \\
\hline
$\Omega h^2$& 0.108 & 0.101 &   \\
\hline  
\end{tabular}} 
\caption{Sample spectra. All masses are in GeV, except for $m_{3/2}$ (in TeV).} 
\label{tab} 
\end{table} 
\end{center}

\subsection{Dependence on $\boldsymbol{\tan\beta}$ and $\boldsymbol{\xi_i}$}

\begin{figure}[h!]
\vspace*{-.4cm}
    \begin{center}
	\hskip -.3cm
       \epsfig{file=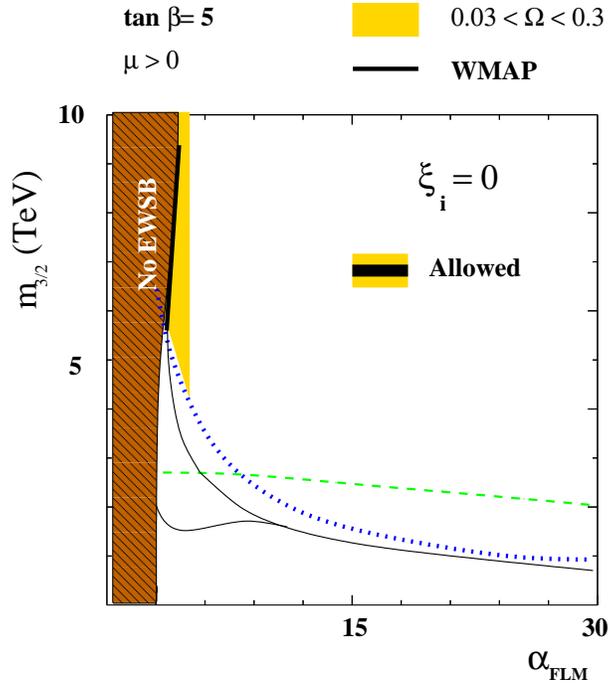,width=0.5\textwidth}
	\vskip -0.1cm
\caption{{\footnotesize
The same as in Fig. \ref{fig:scantb35} but with tan$\beta=5$.} }
        \label{fig:scantb5}
    \end{center}
\vspace*{-.5cm}
\end{figure}

\begin{figure}[h!]
\vspace*{-.4cm}
    \begin{center}
	\hskip -.3cm
       \epsfig{file=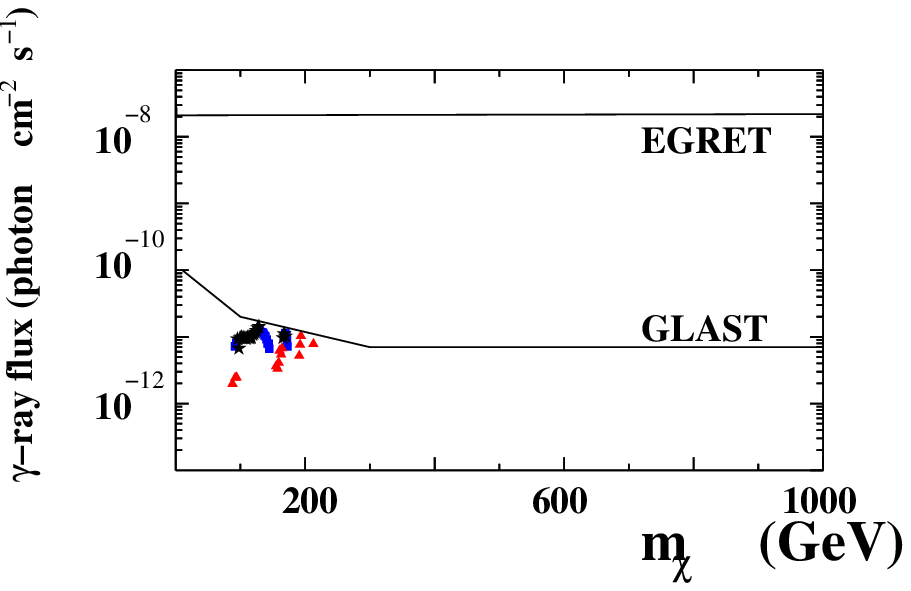,width=0.5\textwidth}
       \epsfig{file=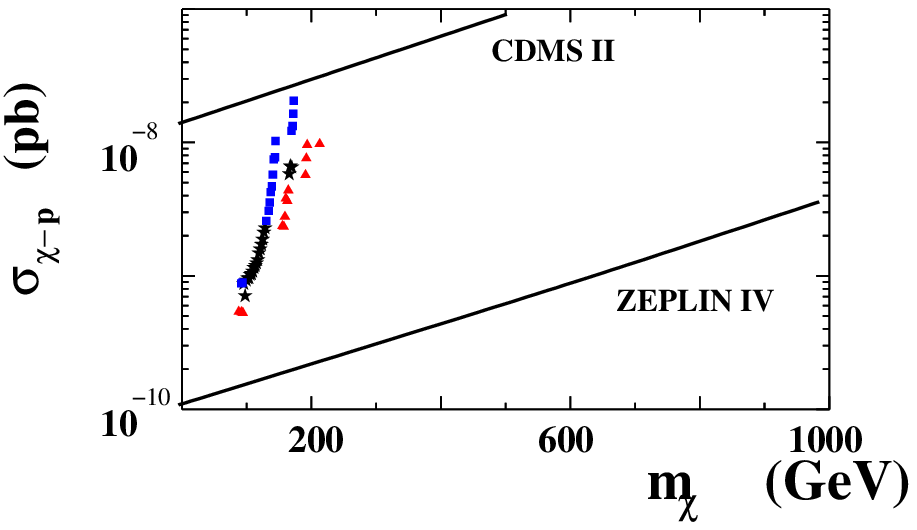,width=0.5\textwidth}

	\vskip -0.1cm
\caption{{\footnotesize 
The same as in Fig. \ref{fig:detectb35} but for tan$\beta=5$.} }
        \label{fig:detectb5}
    \end{center}
\vspace*{-.5cm}
\end{figure}

For lower $\tan\beta$, the ``no EWSB'' constraint relaxes, as explained above.
The allowed region is again on the edge of the ``no EWSB'' area (Fig.~\ref{fig:scantb5}).
An interesting feature, absent in mSUGRA, is that 
$M_1 (M_Z) \simeq M_2(M_Z)$ is possible. In this case, strong   
coannihilation of bino--neutralinos with wino--charginos ($m_{\tilde{\chi}_1^0}\sim m_{\tilde{\chi}_1^+}$)
gives the relic LSP density consistent with WMAP.
On the other hand, the indirect and direct detection rates are somewhat lower
(Fig.~\ref{fig:detectb5}). The neutralinos and charginos are light ($\lsim 200$ GeV)
and can be produced in collider experiments.

\begin{figure}
\vspace*{-.4cm}
    \begin{center}
	\hskip -.3cm
       \epsfig{file=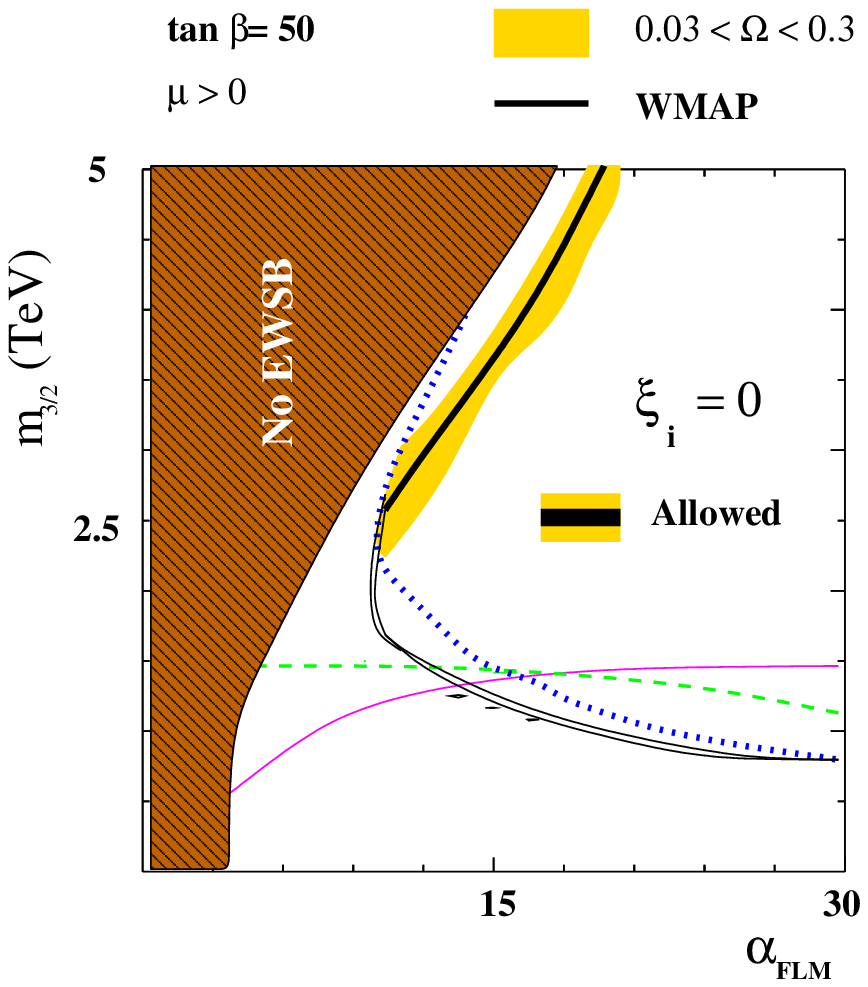,width=0.5\textwidth}
	\vskip -0.1cm
\caption{{\footnotesize
The same as in Fig. \ref{fig:scantb35} but for tan$\beta=50$.} }
        \label{fig:scantb50}
    \end{center}
\vspace*{-.5cm}
\end{figure}

\begin{figure}
\vspace*{-.4cm}
    \begin{center}
	\hskip -.3cm
       \epsfig{file=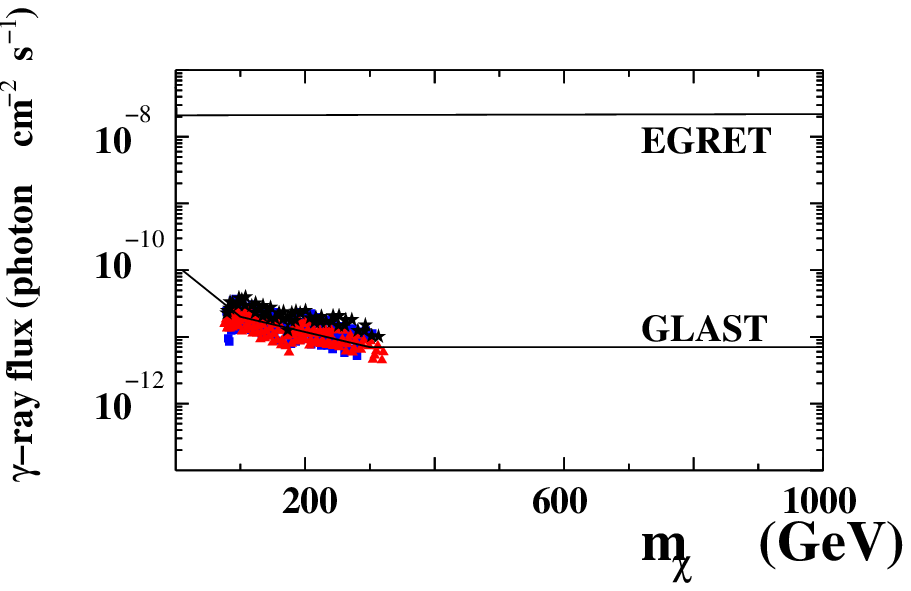,width=0.5\textwidth}
       \epsfig{file=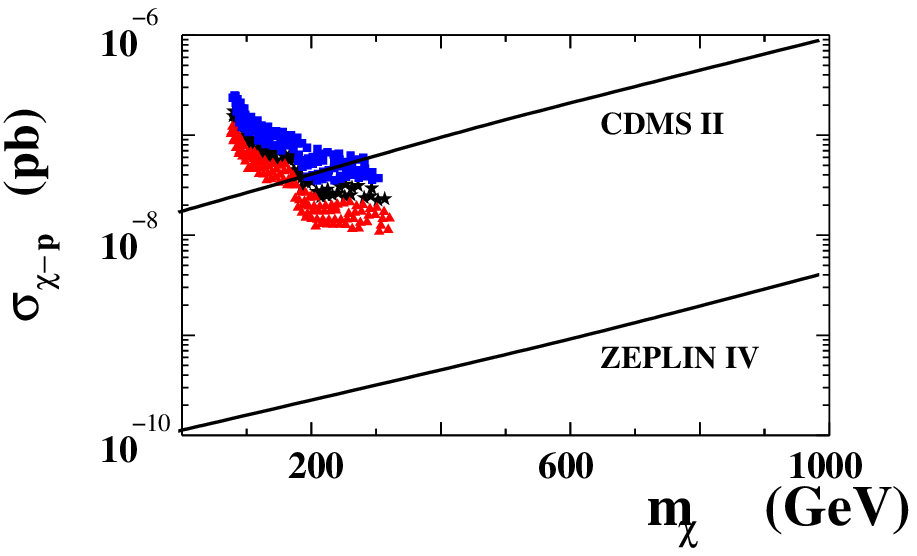,width=0.5\textwidth}

	\vskip -0.1cm
\caption{{\footnotesize 
The same as in Fig. \ref{fig:detectb35} but for tan$\beta=50$.} }
        \label{fig:detectb50}
    \end{center}
\vspace*{-.5cm}
\end{figure}

At $\tan\beta \sim 50$, the picture is similar to the 
$\tan\beta =35$ case (Fig.~\ref{fig:scantb50})
except the detection rates are now enhanced 
(Fig.~\ref{fig:detectb50}). We do not observe the A--pole funnel
for dark matter annihilation since the scalars are too heavy
in the considered parameter space.

\begin{figure}[h!]
\vspace*{-.4cm}
    \begin{center}
	\hskip -.3cm
       \epsfig{file=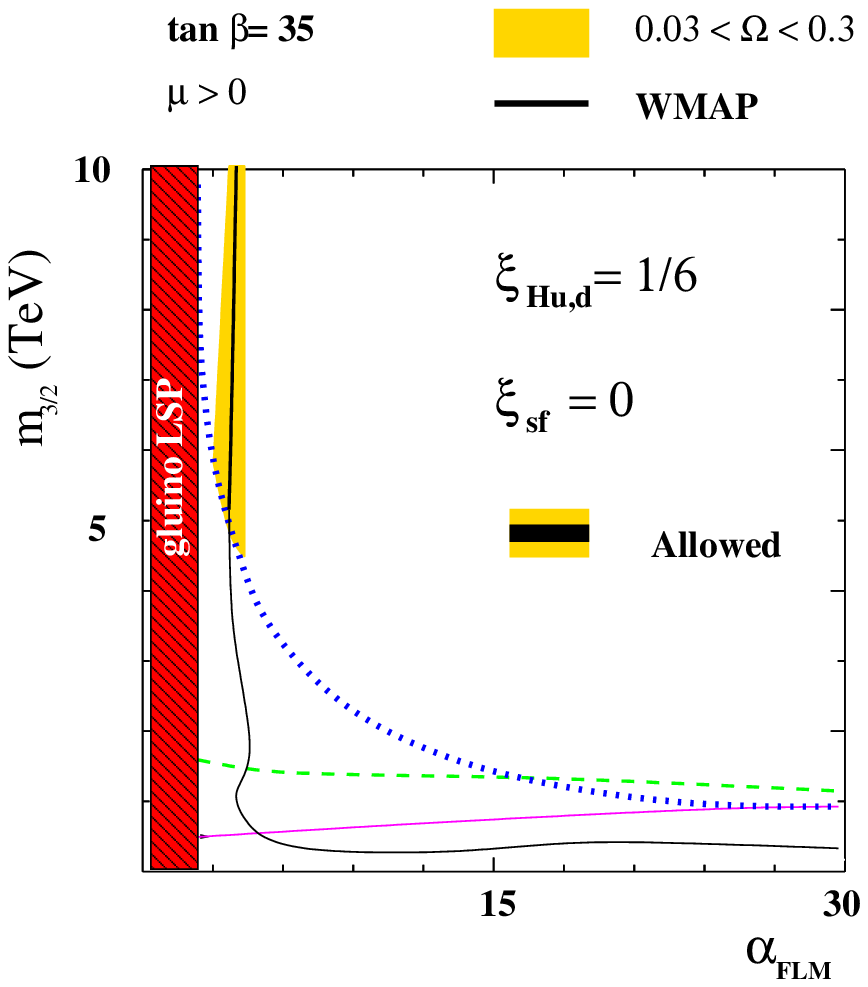,width=0.5\textwidth}
	\vskip -0.1cm
\caption{{\footnotesize
The same as in Fig. \ref{fig:scantb35} but for $\xi_{H_i}=1/6$,
 $\xi_{\tilde f}=0$.} }
        \label{fig:scantb35xihu}
    \end{center}
\vspace*{-.5cm}
\end{figure}

\begin{figure}[h!]
\vspace*{-.4cm}
    \begin{center}
	\hskip -.3cm
       \epsfig{file=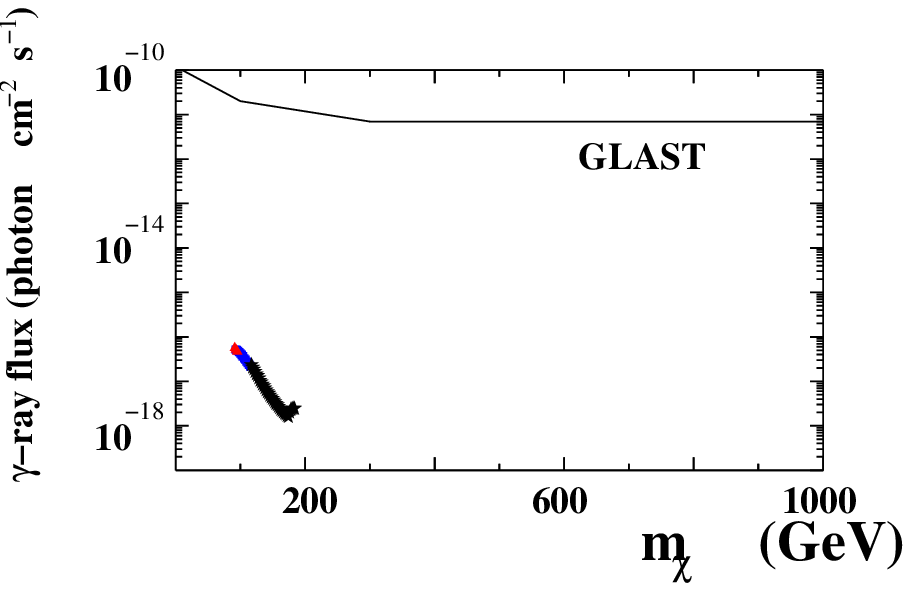,width=0.5\textwidth}
       \epsfig{file=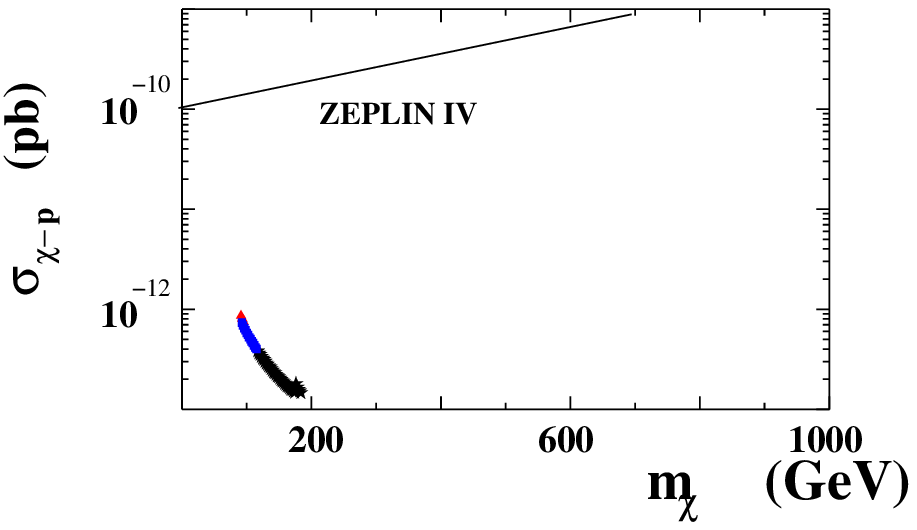,width=0.5\textwidth}

	\vskip -0.1cm
\caption{{\footnotesize 
The same as in Fig. \ref{fig:detectb35} but for $\xi_{H_i}=1/6$,
 $\xi_{\tilde f}=0$.} }
        \label{fig:detectb35xihu}
    \end{center}
\vspace*{-.5cm}
\end{figure}

Increasing $\xi_{H_{u,d}}$ eliminates the ``no EWSB'' 
region (Fig.~\ref{fig:scantb35xihu}), as is
clear from Eq.(\ref{mz}). In this case, a gluino LSP region appears
at small $\alpha_\mathrm{FLM}$. Now the WMAP constraint is satisfied for
larger $\mu$ and the LSP is a bino--wino. Consequently,
the detection rates are suppressed  
(Fig.~\ref{fig:detectb35xihu}).

\begin{figure}[h!]
\vspace*{-.4cm}
    \begin{center}
	\hskip -.3cm
       \epsfig{file=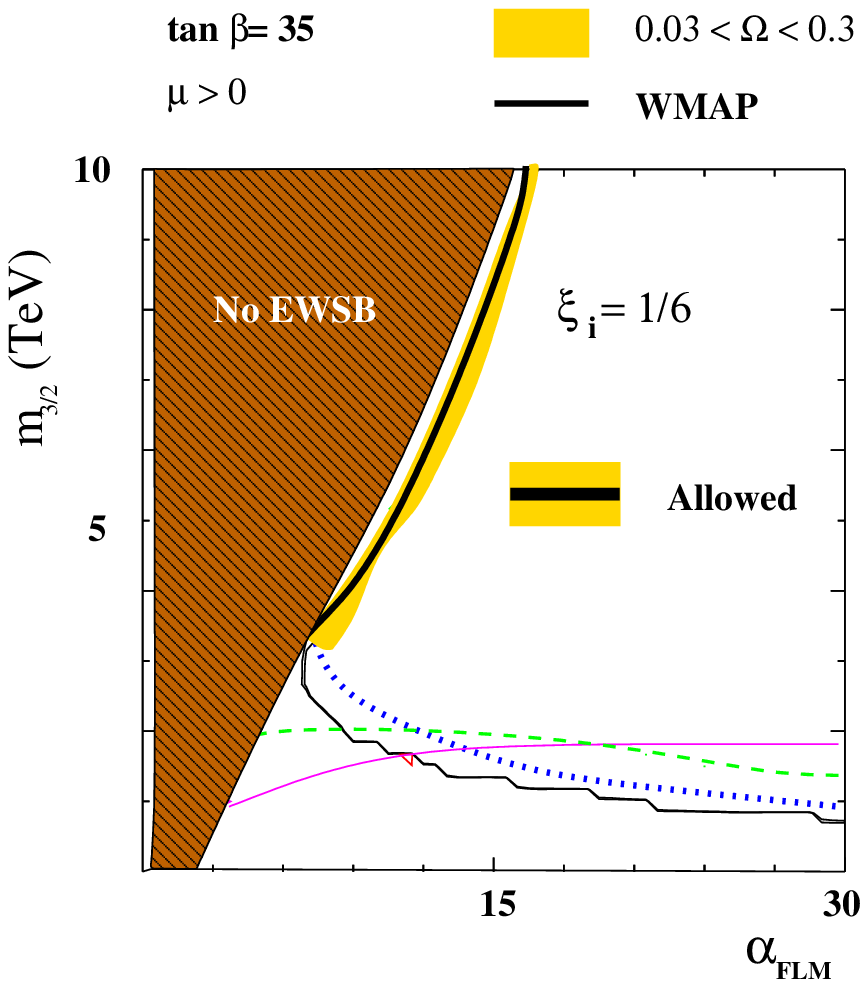,width=0.5\textwidth}
	\vskip -0.1cm
\caption{{\footnotesize
The same as in Fig. \ref{fig:scantb35} but for $\xi_{i}=1/6$.} }
        \label{fig:scantb35xi0.16}
    \end{center}
\vspace*{-.5cm}
\end{figure}

\begin{figure}[h!]
\vspace*{-.4cm}
    \begin{center}
	\hskip -.3cm
       \epsfig{file=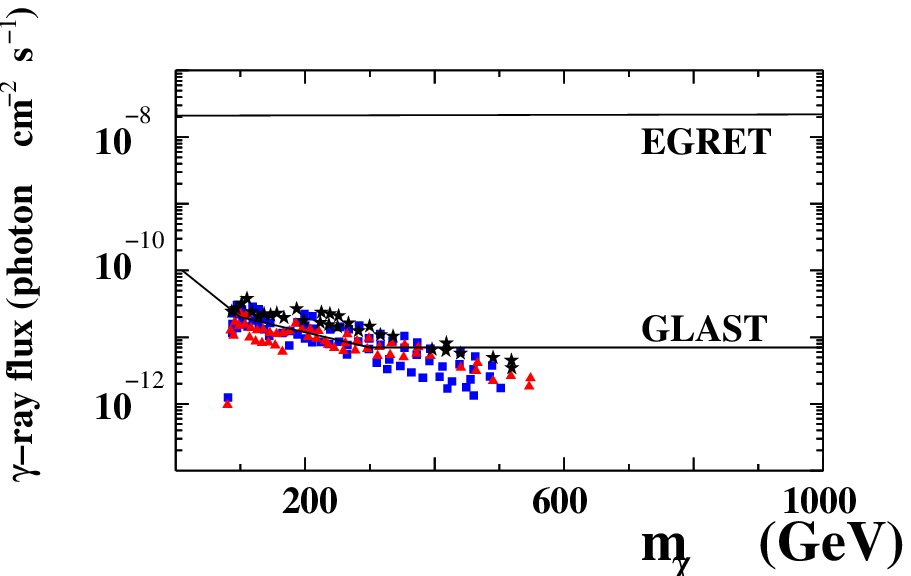,width=0.5\textwidth}
       \epsfig{file=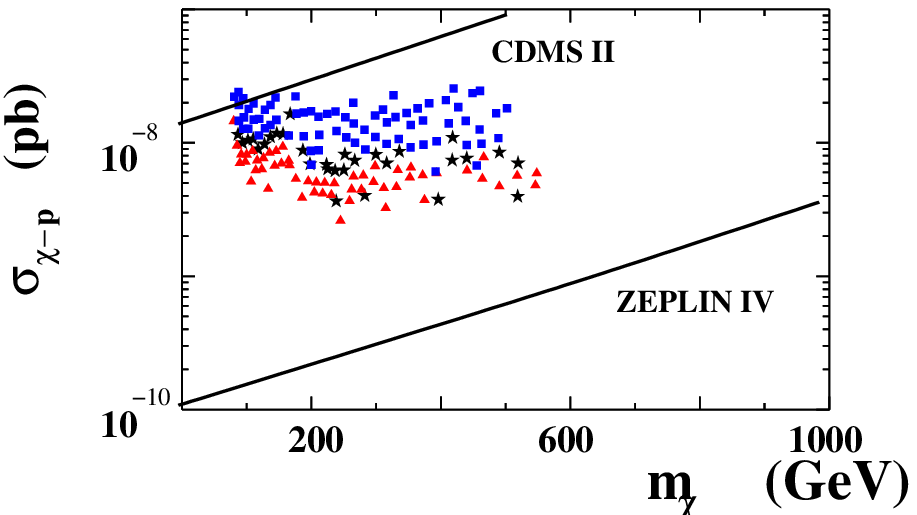,width=0.5\textwidth}

	\vskip -0.1cm
\caption{{\footnotesize 
The same as in Fig. \ref{fig:detectb35} but for $\xi_{i}=1/6$.} }
        \label{fig:detectb35xi0.16}
    \end{center}
\vspace*{-.5cm}
\end{figure}

For all $\xi_i=1/6$, we essentially recover the plots for $\xi_i=0$
and the same conclusions (Figs.~\ref{fig:scantb35xi0.16},\ref{fig:detectb35xi0.16}).

\begin{figure}[h!]
\vspace*{-.4cm}
    \begin{center}
	\hskip -.3cm
       \epsfig{file=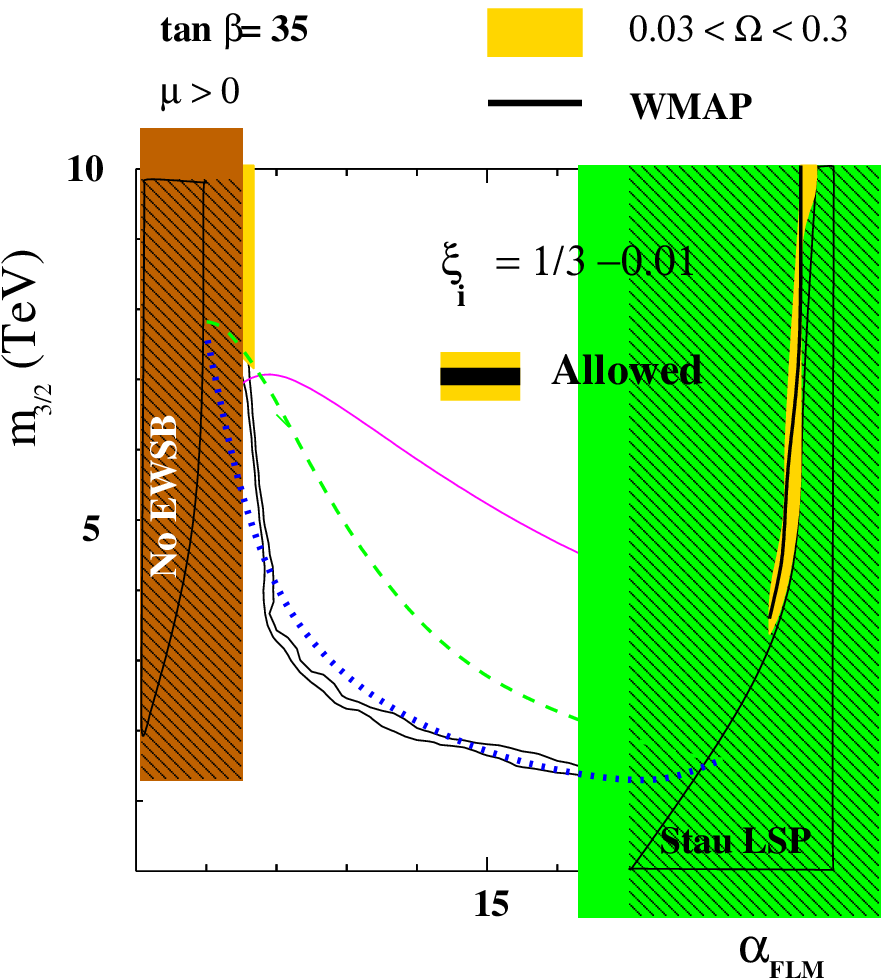,width=0.5\textwidth}
	\vskip -0.1cm
\caption{{\footnotesize
The same as in Fig. \ref{fig:scantb35} but for $\xi_{i}=1/3-10^{-2}$.} }
        \label{fig:scantb35xi0.33}
    \end{center}
\vspace*{-.5cm}
\end{figure}

\begin{figure}[h!]
\vspace*{-.4cm}
    \begin{center}
	\hskip -.3cm
       \epsfig{file=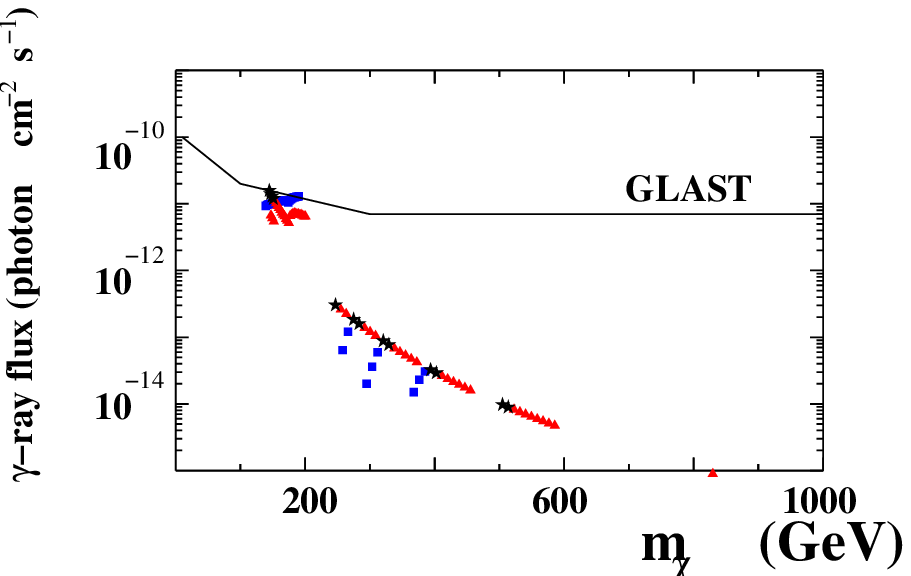,width=0.5\textwidth}
       \epsfig{file=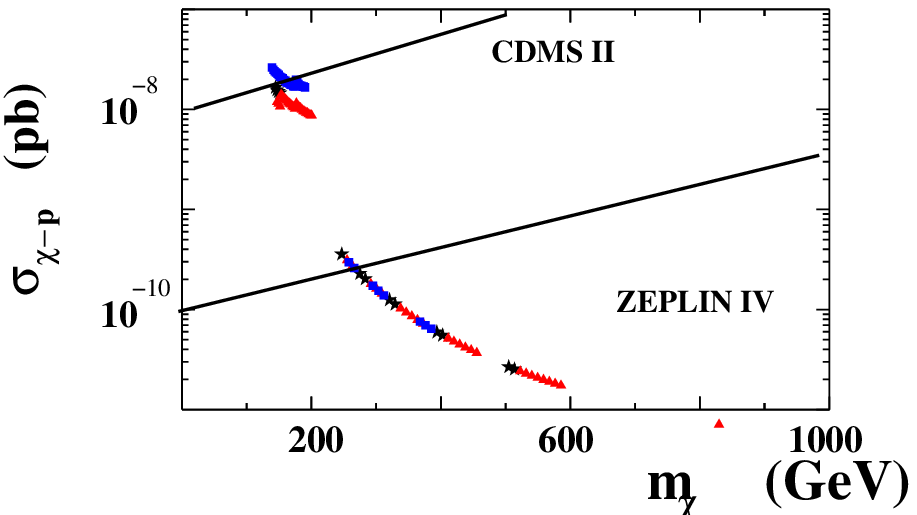,width=0.5\textwidth}

	\vskip -0.1cm
\caption{{\footnotesize 
The same as in Fig. \ref{fig:detectb35} but for $\xi_{i}= 1/3-10^{-2} $.} }
        \label{fig:detectb35xi0.33}
    \end{center}
\vspace*{-.5cm}
\end{figure}

Making the scalars lighter, $\xi_i=1/3-10^{-2}$, changes the picture
dramatically (Fig.~\ref{fig:scantb35xi0.33}). 
The stau can be the LSP, similarly to the mSUGRA case.
Close to the stau LSP  region, $\tilde{\chi}^0_1 \tilde \tau_1$ coannihilation
is efficient and allows for an extra band in the parameter space 
consistent with WMAP. In this region,   $\tilde{\chi}^0_1$ is mainly a bino
and the detection rates are suppressed (Fig.~\ref{fig:detectb35xi0.33}).
The points with significant detection rates correspond to the band
on the left hand side of Fig.~\ref{fig:scantb35xi0.33}, in which
case  $\tilde{\chi}^0_1$ is a mixed higgsino--bino.

For  $\xi_i=1/3$, we recover the ``mirage mediation'' soft terms \cite{Choi:2005ge},
in which case the  (suppressed)  anomaly and modulus contributions to the scalar
masses have to be included. The  corresponding parameter space analysis can be found in 
\cite{Falkowski:2005ck,Baer:2006tb}.

\subsection{Summary}

The ``matter domination'' scenario differs from the ``mirage mediation'' scheme
and mSUGRA in  several phenomenological aspects. First, the scalars are usually
much heavier than the gauginos. This exacerbates the  MSSM  finetuning problem
on one hand, but reduces the  finetuning needed to suppress excessive CP
violation and FCNC, on the other hand\footnote{In our phenomenological study, we have have assumed that  $\xi_i$
and thus the scalar masses are generation--independent. In a more general situation,
this may not be the case and there is a danger of excessive FCNC. However, these
effects are suppressed  (but not completely eliminated)   due to multi--TeV scalar masses. }.  
The gauginos and higgsinos are typically
quite light and accessible to collider searches.  The neutralino dark matter 
can also be detected via the gamma ray flux from the Galactic center as well as
elastic scattering on nuclei.

The typical values of $\alpha_\mathrm{FLM}$ increase compared to  mirage
mediation due to the electroweak symmetry breaking constraint. Also there are no
charged or coloured  tachyons. Non--universality in gaugino masses allows for
$M_1 (M_Z) \simeq M_2(M_Z)$, which is not possible in mSUGRA and leads to
efficient chargino--neutralino coannihilation.

\section{Comments on cosmological problems}

The class  of models we study  do not offer an immediate solution to the
gravitino or moduli problems \cite{Coughlan:1985hh}. The main point of these problems is that
late decaying particles like gravitinos and moduli spoil 
the standard nucleosynthesis (BBN), which has proven to be very successful.
One way to avoid these problems is to make  gravitinos and moduli
sufficiently heavy, 40 TeV or so, such that they decay before the  BBN.
This is possible in our framework for the price of increasing
the sfermion masses as well.
We note, however, that the above  estimate is based on the decay width 
\begin{equation}
 \Gamma~\sim~\frac{m_{\mathrm {scalar} }^3}{M^2} \;,
\end{equation}
with $M \sim M_\mathrm{Pl}$. In practice, this identification may be too rough
and, depending on the K\"ahler potential and other factors, $M$ can be
close to $M_{\mathrm {GUT}}$. In this case, the moduli problem is less severe
and would not require a significant increase in the scalar mass.
In such a scenario, the late--time entropy production will change the picture of
dark matter generation (cf.\ \cite{Drees:2006vh}).

A version of the above  problem, the so called ``moduli--induced gravitino problem'', 
was recently pointed out in the context of the KKLT model with 
the  anti-D3 brane   uplifting  \cite{Endo:2006zj,Nakamura:2006uc}.
In this setup, supersymmetry is broken explicitly and  $m_T\gg m_{3/2}$.
As a result, the branching ratio for the $T$ decays into gravitinos is
of order one which leads to abundant gravitino production and severe
cosmological problems. In the context of spontaneously broken 
supergravity, such a problem is usually absent    \cite{Lebedev:2006qq,Dine:2006ii}  
since the uplifting field $\phi$ typically has a mass comparable to $m_{3/2}$,
\begin{equation}
m_\phi \sim {\cal O}(m_{3/2}) \;.
\end{equation}
This is because, unlike $W(T)$, the uplifting superpotential is not
very steep  \cite{Dudas:2006gr,Abe:2006xp,Kallosh:2006dv}.
$\phi$ dominates the energy--density of the Universe at late times,
however its decay into gravitinos is suppressed and the  
``moduli--induced gravitino problem'' is absent.

\section{Conclusions}

Obtaining phenomenologically interesting vacua in  flux compactifications
is a difficult task. One of the problems is that 
the simple models such as the KKLT  predict the existence of a deep AdS vacuum
which then has to be ``uplifted'' to a dS/Minkowski vacuum by some mechanism.
Here we have advocated the possibility that such uplifting can be provided by hidden 
matter $F$--terms,  along the lines of our earlier work  \cite{Lebedev:2006qq}.
In this case, vacua  with spontaneously broken supersymmetry, small positive
cosmological constant and hierarchically small $m_{3/2}$ can be obtained.
This procedure leads to 
``matter dominated''  supersymmetry breaking, with the modulus contribution
being suppressed. The resulting soft masses are characterized by light gauginos and heavy
scalars.

We have performed a parameter space analysis in this class of models. There are considerable
portions of parameter space consistent with all of the experimental constraints and
accessible to collider searches. We also find good prospects for direct and indirect detection
of neutralino dark matter in the near future.

\noindent
\textbf{Acknowledgements.} 
This work was partially supported by the
European Union 6th framework program MRTN-CT-2004-503069
``Quest for unification", MRTN-CT-2004-005104 ``ForcesUniverse",
MRTN-CT-2006-035863 ``UniverseNet" and
SFB-Transregio 33 ``The Dark Universe" by Deutsche
Forschungsgemeinschaft (DFG).
The work of Y.M. is sponsored by the PAI program PICASSO under contract
PAI--10825VF.

\clearpage

\clearpage

\clearpage

%

\clearpage


%
%
%
\clearpage



\clearpage


\clearpage


\providecommand{\bysame}{\leavevmode\hbox to3em{\hrulefill}\thinspace}

\end{document}